# Second Newtonian branch of salt-free aqueous xanthan solutions

H. Dakhil[1], D. Auhl[2], A. Wierschem[1]

[1] Institute of Fluid Mechanics, Friedrich-Alexander-Universität Erlangen-Nürnberg (FAU), Erlangen, Germany

[2] Polymer Engineering/Polymer Physics, Berlin Institute of Technology (TU Berlin), Berlin, Germany

**Abstract**

We study aqueous xanthan solutions at shear rates up to about $10^5$ s$^{-1}$. At these shear rates, the salt-free solutions show a second Newtonian branch. Depending on the xanthan concentration, we find two different regimes with scaling laws well-known for the zero viscosity of dilute and semidilute particle solutions: The crossover concentration is considerably higher than in the first Newtonian branch, which can be related to the orientation of the polyelectrolytes. In the second regime, the normal stress differences increase with an exponent of about 1, indicating that the polymer solution behaves like nematic liquid crystals or rigid fiber suspensions. In the first regime, the exponent is smaller suggesting that the polyelectrolytes behave more flexible.

**Keywords** Second Newtonian regime, high shear rates, polymer solution, polyelectrolyte, normal stress differences, xanthan

*Corresponding author: Andreas Wierschem
ORCID iD: 0000-0001-7927-2065
Lehrstuhl für Strömungsmechanik
Friedrich-Alexander-Universität Erlangen-Nürnberg
Cauerstr. 4, D-91058 Erlangen
Phone: +49-9131-85-29566
Fax: +49-9131-85-29503
E-mail: andreas.wierschem@fau.de

## I. INTRODUCTION

The rheological properties of polymer solutions in steady shear flows have been studied extensively for many decades. Very dilute solutions have apparently a constant viscosity independent from shear rate [1, 2]. At higher concentrations, the solutions are shear thinning [1-5] and a second Newtonian branch is expected at higher shear rates [4]. Their viscosity functions are usually well described by Carreau-Yasuda and Cross models [6, 7]. At very high concentrations, the solutions may also exhibit yield stress [4].

Many biopolymers are water-soluble polyelectrolytes. The first Newtonian branch of aqueous polyelectrolyte solutions has been studied intensively over a wide range of concentrations, see for instance [8, 9]. In salt-free xanthan solutions, for instance, Wyatt *et al.* identified concentration ranges from dilute to concentrated by their scaling laws [5, 8]: In the semidilute regime, the zero-shear viscosity follows the Fuoss law. In the entangled regime, it increases with concentration according to a power law with an exponent of 3/2. At higher concentrations, the electrostatic blobs overlap and the polyelectrolytes behave as neutral polymers in good solvents. The experimental data is well described by theoretical scaling laws [10, 11].

While the aforementioned classification refers to the first Newtonian branch, where the polymers are close to thermodynamic equilibrium, it remains unclear how these are affected in the second Newtonian branch, i.e. far away from thermodynamic equilibrium. At the onset of this regime, the shear rate has surpassed the fastest inverse relaxation times relevant in shear flows. The average polymer deformation in the second Newtonian branch is high and is not supposed to change significantly anymore. It is often assumed that in this range the polymers are completely stretched and disentangled [12]. Deoxyribonucleic acid (DNA) studies show, however, that the average polymer extension seems to tend to asymptotic values of less than half of its contour length at high shear rates [13, 14]. As a result of competition between Brownian motion and chain convection, the DNA chains carry out a stretching and tumbling motion [14, 15]. The number of entanglements, in any case, decreases at higher shear rates in the shear-thinning region [16]. Hence, if there are different solution states in the second Newtonian branch, their critical concentrations is expected to differ from the first Newtonian branch.

Although the second Newtonian branch is of interest for the fundamental understanding of polymer systems far from thermodynamic equilibrium as well as for applications like coating, turbulence reduction agents or drilling, there is hardly any data on the second Newtonian branch of polymer solutions. The values for the infinite-shear viscosity of the second Newtonian branch are often obtained without reaching the second Newtonian branch by extrapolating deviations from the power-law behavior according to Carreau-Yasuda or Cross models [17, 18]. Yet, there is no systematic study of the second Newtonian branch in polymer solutions. The main reason seems to be the lack of adequate measurement equipment: The shear rates for the second Newtonian branch are usually out of reach for commercial rotational rheometers, which generally cover shear rates below $10^4$ $s^{-1}$. At the expected low infinite-shear viscosities, inertia becomes important, restricting the applicability to lower shear rates and higher viscosities [19]. Higher shear rates can be reached with high-pressure capillary viscometers. However, the viscometers are usually designed for considerably higher viscosities.

There is not only a lack for data on the infinite-shear viscosity, but also on normal stress differences. They have been detected in xanthan solutions only up to shear rates of about $10^3$ s$^{-1}$ [4, 20]. Furthermore, it remains unclear, what the origin of the normal stresses is and how they develop within the second Newtonian regime.

To study the second Newtonian regime of salt-free aqueous polyelectrolyte solutions, shear rates of $10^5$ s$^{-1}$ must be accessed. To this end, we employ a narrow-gap rotational rheometer. In the parallel-disk configuration, it offers a precision in the alignment of both plates better than ±1 µm for unidirectional shear [19, 21-23]. This configuration enables the measurement of normal forces and henceforth of normal stress differences at gap widths of a few micrometers [19]. With the narrow-gap device, we measure the viscosity and study its dependence on the polyelectrolyte concentration at high shear rates. We also detect the normal force, from which we derive the difference between first and second normal stress differences.

We chose xanthan to study aqueous salt-free polyelectrolyte solutions at high shear rates. Xanthan is a polysaccharide produced by Xanthomonas campestris bacteria. The primary structure consists of a backbone with charged trisaccharide side chains on alternating backbone residues [5]. The backbone is similar to that of cellulose [24]. It is one of the stiffest natural biopolymers with a high resistance to mechanical shear [25]. In pure water, xanthan appears elongated and rather forms single helical structures with a contour length of about 1650 nm and a persistence length of about 420 nm [26]. Its solutions have a wide number of industrial and agricultural applications. They are used in coatings, adhesives, drilling fluids, foods, personal care, oil recovery, and agriculture, for instance, as viscosity modifiers, stabilizers, thickeners, drag reducers, drug delivery agents, flocculants, absorbents, dispersants etc. [5, 24, 27-29]. Due to the great importance of polymer solutions in many diverse systems and applications, a fundamental understanding of their flow behavior is crucial.

The article is organized as follows: The preparation of the solutions and the narrow-gap rheometer are described in Sec. II. The experiments and their results of the experiments are reported in Sec. III and discussed in Sec. IV. Finally, the conclusions are summarized in Sec. V.

## II.  MATERIALS AND METHODS

Aqueous xanthan solutions were prepared from xanthan powder (Sigma Aldrich) and deionized water (Sigma Aldrich). According to the supplier, the deionized water had a conductivity of less than 2 µS/cm. Water and xanthan were used as received. Solutions with xanthan concentrations up to 1 wt.% were prepared and observed to be optically clear. The glassware and tools used for mixing and storage of the polymer solutions was carefully cleaned with either ethanol or acetone (reagent grade), then rinsed with purified deionized water to remove all traces of salt prior to use.

The solutions were prepared by dissolving the xanthan powder in the deionized water. They were stirred with a magnetic stirrer for approximately 1 h before being allowed to rest for approximately 24 h at room temperature. Measurements were made one to six days after preparation of the solutions. During storage, the solutions were kept in a refrigerator at 4°C. We have not observed any change in rheological properties of the solutions during the period of six days nor any turbidity of the samples. Seven days after preparation, the solutions were discarded.

All rheological experiments were performed using a modified UDS 200 rotational rheometer from Physica. Experiments were carried out in a parallel-disk configuration. To study the second Newtonian branch at shear rates up to $10^5$ s$^{-1}$, we worked at gap widths as low as 20 µm. To this end, the disks were aligned perpendicular to the axis of rotation to reduce zero-point errors, which are usually about 25 µm and larger [30-32], down to less than ±1 µm [19]. Therefore, we used a bottom plate made from glass with a diameter of 75 mm and an evenness of $\lambda/4$ (Edmund Optics), where $\lambda$ is the testing wavelength (633 nm); the rotating plate of 50 mm diameter with an evenness of $\lambda/10$ was also from glass (Edmund Optics). The top plate was attached to a measurement head of the rheometer with a diameter of 25 mm. To align the bottom plate perpendicular to the axis of rotation, it was fixed to a tripod, which was mounted on the rheometer table. The tripod was aligned with three micrometer screws and fixed to the rheometer with three screws after adjustment. For disk alignment, the gap width was measured with a customized confocal interferometric sensor (STILSA). The sensor was placed underneath the fixed glass plate on a traverse to measure the gap width at different locations. For details on the disk alignment, we refer to [19].

The experiments were carried out at gap widths with a variation of about ±0.8 µm. The temperature was fixed at 25.4±0.1°C using a hood equipped with a heating foil, which was controlled with an H-Tronic TSM 125 temperature control system. Evaporation was minimized by placing wet tissue inside the hood. The setup is sketched in FIG. 1.

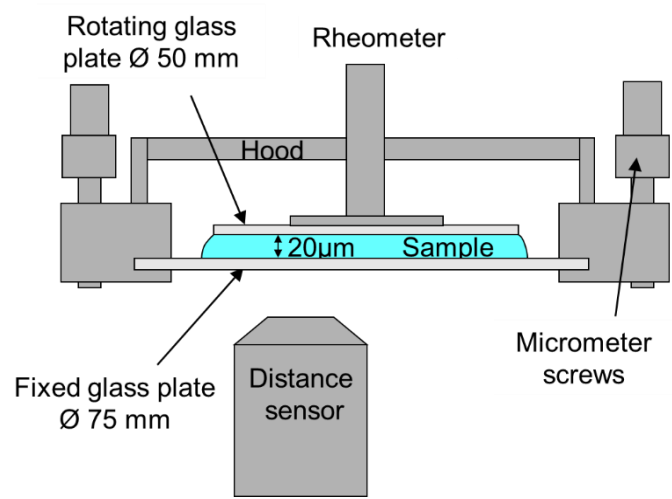

**FIG. 1.** Sketch of the narrow-gap rheometer with homemade temperature control. The gap width between the rheometer plates was measured with a confocal interferometric sensor.

## III. EXPERIMENTAL RESULTS

**VISCOSITY**

FIG. 2 shows viscosity functions of aqueous solutions with xanthan concentrations of 1 wt.%, 0.3 wt.% and 0.1 wt.% that result from experimental runs at different gap widths. For each concentration, the sample was measured first at a gap width of 1000 µm, then at 50 µm and finally at 20 µm. At a gap width of 1000 µm, the samples were pre-sheared at 1000 $s^{-1}$ for 30 s. After 5 min of rest, the normal force was re-set to zero. After another rest of 5 min, the experimental run was carried out by increasing the shear rate stepwise with a measurement time of 10 s per data point. Once the measurements at a gap width of 1000 µm were finished, the gap was lowered to 50 µm and after a waiting time of about 10 min, the run at the new gap width was carried out. For the gap width of 20 µm, this procedure was repeated. In the current study we did not focus on the zero-shear viscosity as its concentration dependence was studied in detail by Wyatt *et al.* [8]. This would demand much longer measurement times [33]. Furthermore, in FIG. 2 we omitted data initially taken below shear rates of 0.1 $s^{-1}$ and data for a concentration of 0.1 wt.% at a gap width of 1000 µm that resulted in torques below 20 times the minimum nominal torque, which is considered a realistic low-torque limit [34, 35].

By lowering the gap width to 50 µm and 20 µm, the maximum shear rate limited by the rotational speed and radial migration [19] shifts to beyond $10^4$ $s^{-1}$ and $10^5$ $s^{-1}$, respectively. Within these limits, the impact of sample inertia on the measured viscosity is negligible [36]. We also checked for viscous heating, which could result in lowering the viscosity, by measuring the disk-surface temperature on the gap side after the experimental runs. Maintaining the shear rate at $10^4$ $s^{-1}$ for 100 s did not show any deviation from ambient temperature; increasing the shear rate to $10^5$ s$^{-1}$ for more than 300 s resulted in a temperature increase of 0.2°C for the highest xanthan concentrations. As seen in FIG. 2, the data at the different gap widths of 1000 µm, 50 µm and 20 µm overlap well. This also indicates that wall slip is not relevant in the considered shear rate region despite using plates with smooth surfaces.

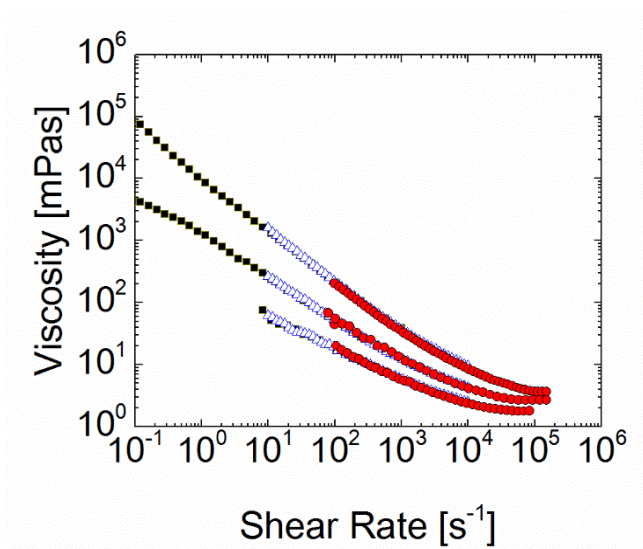

**FIG. 2.** Viscosity functions of aqueous solutions for xanthan concentrations of 0.1 wt.%, 0.3 wt.% and 1 wt.% at different gap widths. Measurements at gap widths of 1000 µm, 50 µm and 20 µm are indicated by solid (black) squares, open (blue) triangles and solid (red) circles, respectively.

After strong shear thinning, the viscosity levels off at high shear rates and a plateau is reached at shear rates well beyond $10^4$ s$^{-1}$ indicating a second Newtonian branch. For all concentrations in FIG. 2, we fit the shear-thinning region of the data measured at 50 µm gap width between shear rates of 10 s$^{-1}$ and $3 \cdot 10^2$ s$^{-1}$ with a power law: $\eta = K\dot{\gamma}^n$, where $\eta$ and $\dot{\gamma}$ are viscosity and shear rate, respectively. For 1 wt.%, 0.3 wt.% and 0.1 wt.%, the power-law exponents $n$ are -0.86, -0.70 and -0.52, respectively.

The concentration dependence of the infinite-shear viscosity is shown in FIG. 3. These data points were obtained from viscosity functions (50 data points) measured between shear rates of $10^4$ s$^{-1}$ and $10^5$ s$^{-1}$ at a gap width of 20 µm. These experiments were carried out in the following manner: After pre-shearing the samples at 1000 s$^{-1}$ for 30 s and 10 min of rest, the experimental runs were carried out by increasing the shear rate stepwise with a measurement time of 10 s per data point. The data points are average values in the second Newtonian branch where they deviate by less than 0.01 mPas. The error bars take into account the remaining uncertainty in the gap width due to imperfect parallelism of the plates and a slightly larger gap, which has been monitored with the sensor during experimental runs and which is supposed to be due to normal forces during shear and due to squeezing of the sample while setting up the narrow gap.

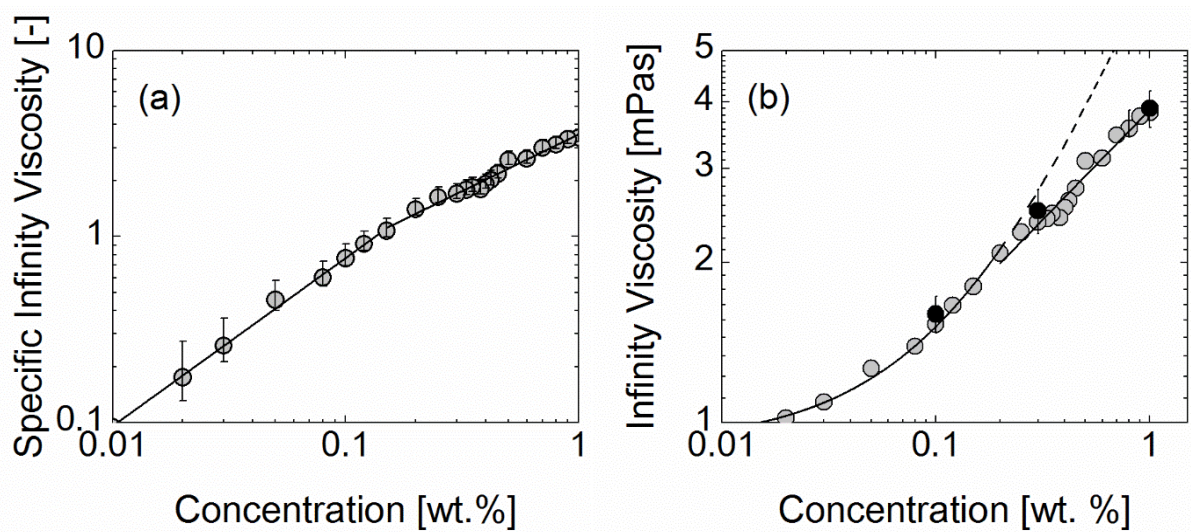

**FIG. 3.** Specific viscosity (a) and viscosity (b) of the second Newtonian branch as a function of the xanthan concentration. Error bars are due to the uncertainty in the gap width for the measurements with the solutions. A representative one is shown in (b). The straight solid lines in (a) are power-law fits to the data; the straight solid line in (b) is a power-law fit, the solid curve in (b) a linear fit, the dashed line an extrapolation of this linear fit. The black circles represent average values from five measurements each according to the procedure shown in FIG. 2. Their error bars indicate the standard deviation between these measurements.

The specific viscosity in FIG. 3(a) was determined with a solvent viscosity of 0.868 mPas, measured with the same equipment as the solution viscosities at a gap width of 20 μm. This value deviates by less than 2.5% from the literature value of 0.889 mPas at 25°C [37]. The diagram reveals that there are two different regimes with a crossover concentration at about 0.15 wt.%. The straight lines are power-law fits to the data. For low concentrations, the power-law exponent of the fit is 0.90±0.04 ($R^2$=0.995) and 0.61±0.03 ($R^2$=0.98) for high concentrations. Diagram (b) shows the viscosity data. Here, the curved line is a linear fit to the viscosity at low concentrations ($R^2$=0.993). The fit value at zero concentration is 0.91 mPas, which differs by about 5% from that measured for the deionized water and by less than 2.5% from the literature value at 25°C [37]. At high concentrations, the straight solid line is a power-law fit. Here, the power-law exponent of the fit is 0.41±0.02 ($R^2$=0.97). The black circles in diagram (b) show the average values of the infinite-shear viscosity from measurements according to the procedure shown in FIG. 2. They fit with the other data within measurement uncertainty.

Characterizing the onset of the second Newtonian plateau by the shear rate where the viscosity is about 10% larger than the infinite-shear viscosity, we find shear rates at about $(6±2) \cdot 10^4$ s$^{-1}$ for the second regime. For the first regime, these shear rates remain at the order of $10^4$ s$^{-1}$ but as a tendency decrease with decreasing concentrations. Extrapolating the power laws from FIG. 2, we find a crossover with the infinite-shear viscosity at shear rates of about $10^4$ s$^{-1}$.

**NORMAL STRESS DIFFERENCES**

During the experimental runs like those shown in FIG. 2 we also detected the normal force. At larger shear rates, particularly beyond $10^4$ s$^{-1}$, the normal forces deviated from the baseline for the studied concentrations. Centrifugal forces result in a negative normal force $F_{N,inertia}$ on the plates [38]:

$$F_{N,inertia} = -\frac{3\pi}{40}\rho(\dot{\gamma}HR)^2, \tag{1}$$

where ρ, $\dot{\gamma}$, H and R are the liquid density, shear rate, gap width and plate radius, respectively. This equation shows that the centrifugal contribution is greatly reduced by narrowing the gap. We checked its applicability with water. After subtracting the centrifugal-forces contribution according to (1), the remaining normal force does not exceed 0.05 N at the highest shear rates as shows the inset of FIG. 4(a).

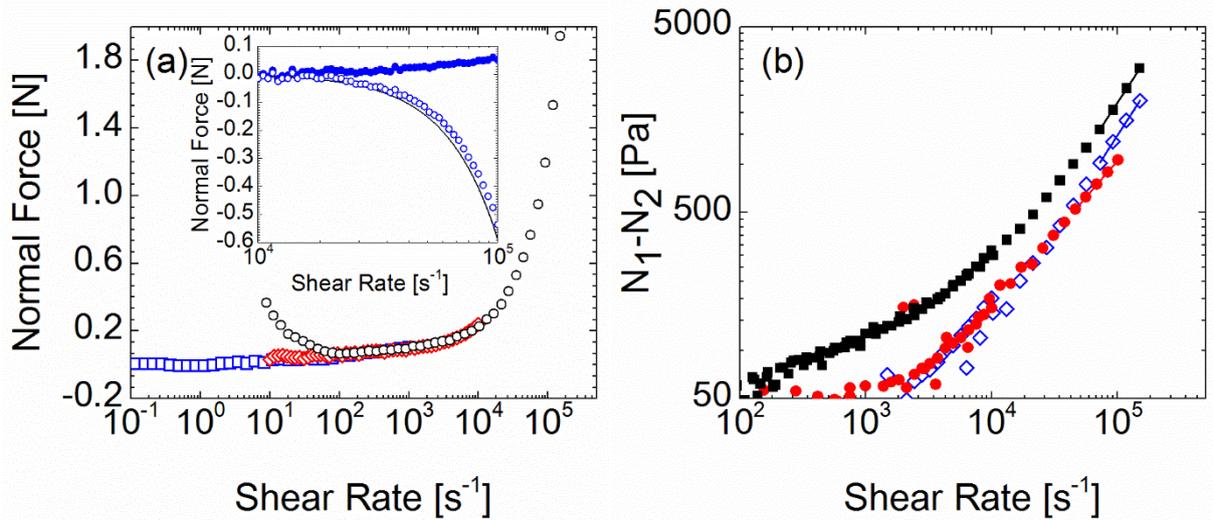

**FIG. 4.** Normal force of a 1 wt.% xanthan solution at gap widths of 1000 μm (□), 50 μm (◊) and 20 μm (○) after baseline and centrifugal force corrections (a) and normal stress difference $N_1$-$N_2$ for 1 wt.% (■), 0.3 wt.% (◊) and 0.1 wt.% (●) xanthan solutions (b). The lines in (b) indicate power-law fits to the data in the second Newtonian branch. Their exponents are 1.04, 1.05 and 0.78 for 1 wt.%, 0.3 wt.% and 0.1 wt.% solutions, respectively. The inset in (a) shows the normal force measured for water at a gap width of 20 μm: Open and solid circles indicate the normal force after baseline correction and after further correcting for centrifugal forces, respectively. The solid line in the inset shows the contribution of centrifugal forces according to (1).

For high xanthan concentrations, we found that even 10 min of rest is not enough to completely release the normal forces created during the reduction of the gap width. Yet, the normal force was released to its lowest values after applying shear during the measurement of the first data points of the viscosity functions, see FIG. 4(a).

Furthermore, the narrower the gap the more the data scatters at low normal forces. This makes it difficult to correct precisely for the baseline. After subtracting the centrifugal-force contribution, we carried out the baseline correction by matching normal force measurements at different gap widths. FIG. 4(a) shows exemplarily the normal force due to normal stress differences for a 1 wt.% xanthan solution after centrifugal-force and baseline corrections for three different gap widths. Beyond shear rates of $10^3$ s$^{-1}$, the normal force due to normal stress differences ever stronger increases and reaches values of about 2 N at the highest shear rates studied. We remark, that the normal force that acts on the rheometer axis, i.e. including the centrifugal-force contribution, usually did not exceed 1 N.

In the parallel-disk configuration, the difference between the first and second normal stress differences, $N_1$–$N_2$, can be determined from the normal force on the plates [39]:

$$N_1 - N_2 = \frac{F_N}{\pi R^2}\left(2 + \frac{d\,ln F_N}{d\,ln\dot{\gamma}}\right), \tag{2}$$

where $F_N$ is the normal force after baseline correction and without the centrifugal-force contribution. The slope correction, $d\,ln\,F_N/d\,ln\,\dot{\gamma}$, was obtained by using a second-order polynomial fit to the logarithm of the data. To suppress contributions at the scatter level, we only considered values for $F_N > 0.1 N$, and for the 20 μm gap $\dot{\gamma} > 2500 s^{-1}$. FIG. 4(b) shows $N_1$–$N_2$. Power-law fits to the data in the second Newtonian branch result in exponents close to 1 for the 1 wt.% and 0.3 wt.% xanthan solutions, and about 0.8 for the 0.1 wt.% solution. A study on the impact of the details of the different correction steps showed that the absolute values for $N_1$–$N_2$ may be affected by the order of ten percent, yet the power-law exponents are robust and vary by just a few percent.

## IV. DISCUSSION

As shows FIG. 2, shear rates up to $10^5$ s$^{-1}$ can be reached by decreasing the gap width in the parallel-disk configuration down to 20 µm. The figure also shows that measurements of different gap size overlap well within the common shear-rate range. Average deviations between the data at 20 µm gap width and that at 50 µm are usually less than 7%, which is the maximum uncertainty due to the gap width variations. Hence, due to the high level of disk alignment, zero-point errors, which result in apparent viscosities that decrease as the gap is narrowed [19, 30-32], are drastically reduced.

Fitting the viscosity in the shear-thinning regime with a power law, we find nice agreement for a concentration for 1 wt.% with the literature value of -0.86 [20]. The value for 0.3 wt.% is close to the -0.72 reported for 0.4% [5]. For 0.1 wt.%, our exponent is somewhat larger than the value of -0.61 mentioned in [5], yet deviations may be due to different fitting ranges, which were not explicitly specified in the reference.

Although we studied the second Newtonian branch over a wide range of concentrations, the viscosity varies only by a factor of 4. Nevertheless, two distinct regimes can be identified, which can be fitted with power laws as shows FIG. 3. These scaling laws are well established for the zero-shear viscosity $\eta_0$ of polymer solutions. For salt-free xanthan solutions, Wyatt and Liberatore identified exponents of 1.6, 0.5, 1.5 and 3.75 for dilute, semidilute, semidilute entangled and concentrated solutions with critical concentrations $c_0^* = 70$ ppm, $c_{e0} = 400$ ppm and $c_{D0} = 2000$ ppm [5].

While $\eta_0$ characterizes the solution close to thermodynamic equilibrium, $\eta_\infty$ is a quantity that describes the solution far from this equilibrium. Average polymer deformation and orientation in the second Newtonian branch are supposed to be high and not to change significantly anymore. It is often assumed that in this range polymers are completely disentangled, stretched and strongly aligned with the flow. In this case, one may expect that the viscosity at high shear rates $\eta_\infty$ obeys scaling laws like those encountered for the zero viscosity of dilute and semidilute particle solutions. This is sustained by the observation of the scaling laws in FIG. 3: In this picture, there is a dilute solution at high shear rates up to an overlap concentration $c_\infty^*$ of about 0.15 wt.%. Beyond this concentration the semidilute particle solution obeys the Fuoss law, indicating that there is no entanglement at these high shear rates in the studied concentration range. This interpretation is sustained by the observation that the criterion for salt-free polyelectrolyte solutions and for neutral polymers in good solvents in the first Newtonian branch $\eta_{sp}(c^*) = 1$ [40] holds nicely for the second Newtonian branch, see FIG. 3(a). Accordingly, the viscosity at overlap concentration is about twice the solvent viscosity . As typical for diluted solutions in the first Newtonian branch [40], we find an exponent of the scaling law close to 1 for the dilute solution, indicating little interaction between the polymers. Yet, it differs from the first Newtonian branch for xanthan, where it is rather 1.6 [5]. Another difference is that we find no indication for a brusque increase in viscosity at overlap, which was attributed to stronger repulsion between the polyelectrolytes due to the electrical charges.

At high shear rates, the overlap concentration is about a factor 20 higher than for the first Newtonian branch. On the other hand, the viscosity is about 20 times smaller than at the end of the dilute regime for $\eta_0$ [5]. This may be due to the much smaller cross section of the polyelectrolytes at high shear rates and hence a drastically reduced interaction in shear direction: In salt-free solutions, polyelectrolytes are fully extended [10, 26, 41]. This results in an overlap concentration $c_0^* \sim L^{-3}$ at low shear rates, where L is the length of the chain [10]. At high shear rates, the chains are supposed to be oriented in flow direction. There, one may rather expect an overlap concentration $c_\infty^* \sim L^{-1} d_{eff}^{-2}$, where $d_{eff}$ is an effective electrostatic diameter. Hence, this diameter is somewhat smaller than the persistence length. Another interesting observation is the fact that $c_\infty^*$ at high shear rates is close to $c_{D0}$ at low shear rates. Beyond this concentration, the polyelectrolytes behave like neutral polymers at low shear rates [5, 10]. The fact that we find at the high shear rates a scaling law exponent close to that of the Fuoss law indicates that the overlap is avoided at these concentrations due to the high degree of order.

Normal forces become noticeable beyond $10^3$ s$^{-1}$, as shows FIG. 4. Normal forces after subtraction of the centrifugal-force contribution and thus $N_1$–$N_2$ increase with concentration and with shear rate. In the second Newtonian regime, $N_1$–$N_2$ obeys a power-law dependence on the shear rate with an exponent close to 1 in semidilute particle solutions, i.e. beyond $c_\infty^*$, see FIG. 4(b); at the concentration of 0.1 wt.%, we find an exponent smaller than 1.

The power-law dependence of $N_1$–$N_2$ in the second Newtonian regime is similar to that of suspensions of rodlike particles. Like in our semidilute particle solutions, the normal stress difference of nematic liquid crystals [42] and of semi-concentrated rigid fiber suspensions is proportional to the shear rate [43-45]. The normal stresses in the fiber suspensions are mainly due to collisions between the fibers [45]. For flexible fibers, the linear relationship between normal stress differences and shear rate only holds for the semidilute regime [45]. At higher concentrations, a power-law dependence with an exponent below 1 has been observed [45-47]. Note that the second normal stress difference is much smaller than the first one in fiber suspensions and, hence, the normal stress difference measured with a parallel-disk device can be taken to be $N_1$ [46].

The results on the normal stress difference together with the scaling laws for the concentration dependence of the viscosity suggest the following picture for the second Newtonian branch: The disentangled polyelectrolytes create normal stress differences rather due to collisions than due to viscoelastic deformation. In line with the observation that single xanthan molecules appear elongated in pure water [26], this suggests that shear thinning is rather due to disentanglement and orientation of the xanthan polymers than due to deformation and stretch. In the semidilute particle regime, the linear dependence of the normal stress difference on shear rate indicates that the polyelectrolytes are rather rigid, apparently due to interaction with each other as described by the Fuoss law. Hence, in this regime one may consider them as nematic liquid crystals. In the dilute regime, the polyelectrolytes act more flexible as suggests the exponent lower than 1 for the relation between normal stress differences and shear rate. This is apparently possible since stabilization of the polyelectrolyte form by electrostatic repulsion from surrounding neighbors during collision is too weak.

## V. CONCLUSIONS

We studied the high shear-rate flow behavior of aqueous salt-free xanthan solutions up to shear rates of about $10^5$ $s^{-1}$ and mainly focus on the second Newtonian branch. At shear rates well beyond $10^3$ $s^{-1}$, the viscosity function levels off and reaches a plateau. Depending on the xanthan concentration, we find two different regimes for the second Newtonian branch that we identify by their scaling laws for the viscosity. Up to a concentration of about 0.15 wt.%, we find a linear dependence of the specific viscosity in the second Newtonian branch. Beyond that concentration, the exponent is close to the Fuoss law for the first Newtonian branch, indicating that the polymers are disentangled. At the high shear rates studied, the normal stress differences increase with an exponent of about 1 in the second Newtonian branch, indicating that they are rather due to collisions between the polymers than due to viscoelastic deformation of the polymers. The results indicate that at high concentrations the solutions behave like nematic liquid crystals.

**ACKNOWLEDGMENTS**

The authors are thankful to Mrs. N. Bader, Mr. P. Kremer and Mr. T. Zinnöcker for collaborating in preliminary experimental studies. The support from Deutsche Forschungsgemeinschaft through WI 2672/9-1 is gratefully acknowledged.